\documentclass[twocolumn,showpacs,aps,pra]{revtex4}
\usepackage{epsfig,amssymb,amsmath,amsmath}

\newcommand{\dpsi}{\delta \hat{\Psi}}
\newcommand{\rvec}{\vec{r}\,}
\newcommand{\ha}{\hat{a}}
\newcommand{\hb}{\hat{b}}
\newcommand{\hc}{\hat{c}}
\newcommand{\la}{\langle}
\newcommand{\ra}{\rangle}
\newcommand{\hpsi}{\hat{\Psi}}
\newcommand{\ex}{^{\rm ex}}
\newcommand{\hlam}{\hat{\Lambda}}
\newcommand{\bna}{\bar{N}_a}
\newcommand{\bnb}{\bar{N}_b}
\newcommand{\ddt}{\frac{d}{dt}}
\newcommand{\hna}{\hat{N}_{0a}}
\newcommand{\hnb}{\hat{N}_{0b}}

\newlength{\figwidth}
\begin{document}

%\draft

\title{Bogoliubov theory of entanglement in  a
  Bose-Einstein condensate}

\author{Anders S\o ndberg S\o rensen}
%\email{anderss@ifa.au.dk}
\affiliation{Institute of Physics and Astronomy, University of Aarhus, \\
DK-8000 Aarhus C, Denmark}

\begin{abstract}
We consider a Bose-Einstein condensate which is illuminated by a
short resonant 
light pulse that coherently couples two internal states of the
atoms. We show that  the
subsequent time evolution prepares the
atoms  in an interesting 
entangled state called a spin squeezed state. This evolution is
analysed in
detail by developing a Bogoliubov theory which describes the entanglement of
the atoms. Our calculation is a consistent expansion in $1/\sqrt{N}$, where
$N$ is the number of particles in the condensate, and our theory predict
that it is possible to produce spin squeezing by at least a factor of
$1/\sqrt{N}$. Within the Bogoliubov approximation this result is
independent of temperature.
\end{abstract}


\maketitle

\section{Introduction}
% The physics of quantum entanglement is very interesting both from a
% fundamental and a practical perspective. The use of entangled particles to
% violate the Bell inequalities has increased of knowledge of the physical
% reality \cite{aspect} and the discovery of practical applications of
% entanglement in 
% quantum information \cite{physworld} has recently boosted the interested
% in the field.
One of the main
problems
in the experimental explorations of entanglement is the engineering of
a controlled interaction between 
individual quantum particles.  
%In Bose-Einstein
%condensate in a dilute atomic gas the engineering of such an interaction
%between the particles is absolutely no problem at all. 
Such  an
interaction among 
the particles is automatically present in 
Bose-Einstein condensed dilute atomic gases
\cite{Cornell,Ketterle,General}, where 
the experimental results show clear
signatures of the collisions of the  particles. It is
therefore natural to try to 
exploit this interaction to create entanglement and  a
number  of proposals have appeared which use the collisional interaction to
entangle the atoms in the condensate
\cite{anders-nature,duan,pu,meystre,uffe-posp,uffe2,helmerson,duan2}.  

In Ref.~\cite{anders-nature} we proposed  a realistic experiment where the
atoms in a Bose-Einstein condensate are illuminated by a single resonant
pulse. After this pulse the system evolves freely, and the
collisional interaction among the particles prepares the
atoms in an entangled state. The created entangled state even has
practical applications in atomic clocks, where it can be used to
increase the precision  significantly \cite{win-sq}.   
In \cite{anders-nature} we presented an approximate simulation of the
condensate dynamics, where the state vector was expanded on subspaces
containing different number of atoms, and where the spatial wavefunction in
each subspace was evolved with the
Gross-Pitaevskii equation. This 
calculation  indicated that a substantial entanglement can be
created, but the validity of the ansatz used in the calculation and the
behaviour of the proposal at non-zero temperature is difficult to determine
from the simulation. An exact calculation with a stochastic method
\cite{uffe-posp} has later confirmed that the proposal is indeed capable
of producing substantial entanglement. But due to the noise in this
calculation, it is difficult to estimate the 
exact amount of entanglement from this procedure 
and the method is also 
difficult to apply in the regime where most experiments are currently
operating. 

In this paper we investigate the proposal \cite{anders-nature}
in more detail. We first develop a Bogoliubov theory for a two component
condensate and then we apply it to describe the entanglement of the atoms in
the condensate. Our theory is a consistent expansion in the ratio between
non-condensed and condensed particles and the validity of the approximation
can directly be investigated for a given experimental
configuration. To lowest order, the Bogoliubov theory can
be used 
to describe the effect of a non-vanishing temperature, and
our theory indicates that within the Bogoliubov approximation the
created entanglement 
is independent of temperature. The results of the
Bogoliubov theory agrees very well with the results of the approximate
method used in Ref.\ \cite{anders-nature} and thus confirms the
validity of this method. 

\subsection{Considered experimental setup} 
\label{setup}
We consider a collection of $N$ atoms with two internal states $a$ and $b$. 
We assume that we first create a condensate consisting only of atoms in the
internal state $a$. After the preparation of the condensate
a fast 
resonant 
pulse is applied to the atoms. The pulse is assumed to be much faster than
any other time scale in the problem, so that the only action of the pulse
is to mix the internal states $|a\ra\rightarrow \cos(\phi/2)|a\ra
+\sin(\phi/2)|b\ra$ and  $|b\ra\rightarrow \cos(\phi/2)|b\ra
-\sin(\phi/2)|a\ra$. The theory developed in section \ref{theory} and
\ref{solve} 
applies for all values of the angle $\phi$ but we only go into details
with the case where the resonant interaction is a $\pi/2$ pulse,
$\phi=\pi/2$. In this paper we shall describe our atoms in the second
quantization using field operators $\hpsi_a$ and $\hpsi_b$, and we calculate
the time evolution in the Heisenberg picture, where  
the action of the pulse is
given by
\begin{equation}
\begin{split}
\hat{\Psi}_a(\rvec,t=0^+)=&\cos{\left( \frac{\phi}{2}\right)}
\hat{\Psi}_a(\rvec,t=0^-) 
       \\ 
 &-\sin{\left( \frac{\phi}{2}\right)} 
               \hat{\Psi}_b(\rvec,t=0^-)  \\
\hat{\Psi}_b(\rvec,t=0^+)=&\cos{\left( \frac{\phi}{2}\right)}
\hat{\Psi}_b(\rvec,t=0^-) 
                    \\
 &+\sin{\left( \frac{\phi}{2}\right)} \hat{\Psi}_a(\rvec,t=0^-)  
\end{split}
\label{pulse}
\end{equation}
($t=0^-$ and  $t=0^+$ denotes the time just before and after
the  pulse). After the pulse we let the system
evolve and perform a measurement on it at a later time.

The resonant pulse splits the condensate into an  $a$ and a $b$
component with a well 
defined phase between them; if one applies a new pulse shortly after the
first, it is possible to produce interferences between the two component
and, e.g., transfer all the atoms back into the $a$ state. As the two
components evolve with time there is nothing to maintain the relative
phase between them, and  the
relative phase will spread in time because each of the two components
do not have 
a well defined number of particles
\cite{lewenstein,sinatra,castin-dum}. Due to the spreading of the
relative 
phase, the magnitude of interference terms like $\la \hpsi_a^\dagger
\hpsi_b\ra$ will decrease with time and it  will no longer be possible to
see interferences between the two components. This process is referred to
as phase collapse. 

Instead of focusing on the interferences between the condensates it is
fruitful to consider other observables of the system. By considering
fluctuations in the particle numbers one can show that the time evolution
prepares the atoms in an entangled state called a spin squeezed stated. 

\subsection{Spin squeezing}
\label{squeez-intro}
Suppose that we prepare $N$ atoms in an equal superposition of two
internal state 
$(|a\ra +|b\ra)/\sqrt{2}$  
and measure the number of particles in the $a$ state.  Because the
particles are 
independent of each other the result of the measurement will be a
distributed according to the binomial distribution with a 
mean value $\bna=N/2$ and a variance $N/4$. A spin squeezed state is a
state where the 
atoms are prepared in a suitable entangled state such that the
variance (or noise) of 
this measurement is reduced. 

A more convenient way to represent the atoms is
to consider each atom as a spin-$1/2$ particle with the internal states
$|a\ra$ and $|b\ra$ representing the spin up and spin down states
respectively. In this language the noise in the counting statistics is
represented by the variance of the $J_z$ operator $(\Delta J_z)^2$, where
the collective $J_x$, $J_y$, and $J_z$ operators are obtained by summing
the spin operators for the individual atoms. Expressed in terms of the
spin operators, spin
squeezing is  the reduction of the noise of $J_z$ (or any other spin
component).  Recently the first experimental realizations of such spin 
squeezed states have been achieved \cite{polzik1,bigelow,kasevich,polzik2}.

In Ramsey spectroscopy, as used for instance in atomic clocks, a
signal is recorded which is 
proportional to the length of the spin in a given direction, say the $x$
direction, and the noise of the signal is proportional to the noise of a
spin component $J_\theta=\cos(\theta)J_z+\sin(\theta)J_y$ perpendicular to
the mean spin. 
The precision of atomic clocks is currently limited by the spin noise
$(\Delta J_\theta)^2$ \cite{santarelli}, and by using a spin
squeezed state where this noise is smaller than in the unsqueezed
states, which are used today,  it is
possible to increase the precision of the clock.
Wineland {\it et al.} \cite{win-sq} have analysed this possibility 
and have 
shown that the frequency variance can
be reduced by a factor of
\begin{equation}
\xi_\theta^2=\frac{N(\Delta J_\theta)^2}{\la J_x\ra^2}
\label{defxi}
\end{equation}
by preparing the atoms in a spin squeezed state with
$\xi_\theta^2<1$. This points to an 
interesting application of spin squeezed states in atomic clocks. 

It has also been shown that a reduction of the squeezing parameter below
unity 
$\xi_\theta^2<1$ requires the atoms to be in an entangled state
\cite{anders-nature}, and to quantify the entanglement which may
be obtained in the considered experimental setup, we shall
determine the reduction of the squeezing parameter $\xi^2_\theta$.  
A different characterization of the squeezing is
presented in Ref.\ \cite{anders-maxsqueez}, where the depth of entanglement in
a collection of atoms is identified. 

To see that the proposed
experimental setup produces squeezing we shall first consider the short
time evolution. At time $t=0^+$ we
have $\xi_\theta^2=1$ for all $\theta$, and to prove that squeezing is
produced, it is sufficient to show that the
time derivative 
of $\xi_\theta^2$ is negative for a suitable choice of $\theta$.

In the limit of very low temperatures a collection of atoms with two internal
states $a$ and $b$ are described by the second quantized Hamiltonian 
\begin{equation}
\begin{split}
H=&\sum_{j=a,b} \int d^3 r \hat{\Psi}_j^\dagger(\rvec) H_{0,j}
\hat{\Psi}_j(\rvec) \\
& +\frac{1}{2} \sum_{j=a,b} g_{jj} \int d^3 r \hat{\Psi}_j^\dagger(\rvec) 
 \hat{\Psi}_j^\dagger(\rvec)  \hat{\Psi}_j(\rvec)
  \hat{\Psi}_j(\rvec)  \\
 &+g_{ab}\int d^3 r  \hat{\Psi}_a^\dagger(\rvec)
  \hat{\Psi}_b^\dagger(\rvec)  \hat{\Psi}_a(\rvec)
  \hat{\Psi}_b(\rvec),
\end{split}
\label{hamilton}
\end{equation}
where $H_{0,j}$ is the one particle Hamiltonian for atoms in
state $j$ including the kinetic energy and the external trapping
potential $V_j(\rvec)$, and $g_{jk}=4\pi \hbar^2
a_{s,jk}/M$ is the strength of the interaction between particles
of type $j$ and $k$, expressed in terms of the scattering length
$a_{s,jk}$ and the atomic mass $M$. Here we have assumed that there are
no spin changing collisions. A specific experimental setup where it is
possible to exclude such spin changing collision with sodium atoms in an
optical trap is described in \cite{anders-nature}.

With the Hamiltonian (\ref{hamilton}) we may find the equation of motion
for the field operator $\hpsi_a$ by taking the commutator with the
Hamiltonian (using $\hbar=1$)
\begin{equation}
i\frac{d}{dt} \hpsi_a= H_{0,a} \hpsi_a+g_{aa}\hpsi_a^\dagger \hpsi_a\hpsi_a
          +g_{ab}\hpsi_b^\dagger\hpsi_b\hpsi_a.
\label{dpsidt}
\end{equation}
For brevity most of the equations in this papers will
only be presented for the field operator for the $a$ component. The
corresponding equations for the $b$ component can be achieved
by exchanging subscripts $a$ and $b$. Since we work in the Heisenberg
picture, essentially all operators and functions appearing  in this paper
will depend on time, and for brevity we shall in most cases omit
the time argument on operators and functions. 
Unless specified otherwise operators and functions, like the
field operators $\hpsi_a$ and $\hpsi_b$ in Eq.\ (\ref{dpsidt}) should
always be considered as functions of the time $t$.
From the derivative in (\ref{dpsidt}) we calculate the
time derivative of the squeezing parameter $\xi_\theta^2$ at the time
$t=0^+$, and by using the relation (\ref{pulse}) with $\phi=\pi/2$ we can
express the derivative in terms of the field operators before the pulse 
\begin{equation}
\begin{split}
\ddt \xi_\theta^2=&\sin(2\theta)\, \frac{g_{aa}+g_{bb}-2g_{ab}}{2N} \times
  \\
& \ \   \int d^3r \la \hpsi_a^\dagger(\rvec,t=0^-)^2
 \hpsi_a(\rvec,t=0^-)^2\ra.
\end{split}
\label{dxidt}
\end{equation}
If $g_{aa}+g_{bb}\neq 2g_{ab}$ we can always choose the angle $\theta$
such that the expression on the 
right hand side is negative and this {\it exact} equation immediately
proves that 
squeezing will be produced. The time derivative
is proportional to the two point correlation function which in the limit of
an ideal condensate is $N(N-1)\int d^3r |\Phi|^4$, where $\Phi$ is the
condensate wavefunction before the pulse. 

The above argument shows that squeezing will be produced but it does
not quantify the amount of squeezing which is produced. A simple
estimate of the 
squeezing can be obtained by making a single mode approximation to
each of the components. In
this approximation the part of the Hamiltonian (\ref{hamilton}) which is
responsible for spin squeezing can be written as
\cite{anders-nature,meystre,uffe-posp} 
\begin{equation}
H_{\rm spin}=\chi J_z^2
\label{hspin}
\end{equation}
with the coupling constant $\chi$ given by
\begin{equation}
\chi(t)=\int d^3
r \frac{g_{aa}|\Phi_a\ex|^4+ g_{bb}|\Phi_b\ex|^4-
2g_{ab}|\Phi_a\ex|^2|\Phi_b\ex|^2}{2},  
\label{chisimple}
\end{equation}
where $\Phi_a\ex$ ($\Phi_b\ex$) is the condensate wavefunction for
particles of type $a$ ($b$) (the ``${\rm ex}$'' is used to indicate
the exact wavefunction 
because it will later be important to distinguish it
from low order approximations to it).

The squeezing arising from the Hamiltonian (\ref{hspin}) can be solved
analytically \cite{ueda}. The obtainable squeezing is approximately
$\xi^2_\theta \approx N^{-2/3}$, and this indicate that strong
squeezing can be produced if the condensate contains a large number of
atoms. The
validity of the 
arguments leading to the Hamiltonian (\ref{hspin}) are not quite clear from
the derivation, and it is the main purpose of this paper to investigate the
validity of this simple description. As we
shall see below, the Hamiltonian (\ref{hspin}) is indeed a good
approximation to the dynamics of the system but the coupling constant
(\ref{chisimple}) has to be modified slightly. To obtain this result
we describe the  system by 
Bogoliubov theory.

\subsection{Bogoliubov theory}

The Hamiltonian (\ref{hamilton}) and the resulting Eq.~(\ref{dpsidt}) are far
to  complicated to be solved in
general and approximations have to be applied. Here we shall determine
the time evolution from the Hamiltonian (\ref{hamilton}) by applying the
Bogoliubov approximation. The standard formulation of the Bogoliubov
approximation assumes a broken symmetry such that the field operators have
a non-vanishing mean value $\langle
\hat{\Psi}_a(\rvec,t)\rangle=\sqrt{\bna}\Phi_a\ex(\rvec,t)$,  where $\bna$ 
is the mean number of particles of type
$a$. The Bogoliubov method proceeds by
splitting the field operators according to
\begin{equation}
\begin{split}
 \hat{\Psi}_a(\rvec,t)&=\sqrt{N_a}\Phi_a\ex(\rvec,t) +\dpsi_a(\rvec,t)
  \\
 \hat{\Psi}_b(\rvec,t)&=\sqrt{N_b}\Phi_b\ex(\rvec,t) +\dpsi_b(\rvec,t) .
\label{symbreak}
\end{split}
\end{equation}
The idea behind this splitting
is that the fluctuations $\dpsi_a$ and $\dpsi_b$
are much smaller than the contribution of the condensates, and simple
equations for the field operators are obtained by expanding the
Hamiltonian (\ref{hamilton}) to  low order in the $\dpsi$s. 
% As a consequence of the breaking of the symmetry and the omission of the
% terms of higher order in the $\dpsi$s, the Bogoliubov method does not conserve
% the  
% total number of particles, but this is of minor importance as long as
% the fluctuations in the number of 
% particles is much lower than the total number.

The splitting (\ref{symbreak}) gives some nice properties, such as simple
commutation relations for the $\dpsi$ operators, but unfortunately it is not
well suited for our 
present purpose. As we shall see below, the phase collapse and entanglement
mainly arise from the evolution of the atoms in the condensate modes. In
Eq. (\ref{symbreak})  the
quantum fluctuations of the condensate modes and fluctuations perpendicular
to these modes are treated equally, and this is not a good
approximation to the  system.  In a direct numerical
integration we have found that the $\dpsi$s grow
very quickly due to the dynamics in 
the condensate modes so that the assumptions of the theory break
down after a short period of time \cite{rs}.

To circumvent the problems associated with the splitting (\ref{symbreak})
we employ a different splitting which enables a better description of the
evolution in the condensate modes 
\begin{equation}
\begin{split}
 \hat{\Psi}_a(\rvec,t)&=\ha\Phi_{a}\ex(\rvec,t) +\dpsi_a(\rvec,t)
 \\
 \hat{\Psi}_b(\rvec,t)&=\hb\Phi_{b}\ex(\rvec,t) +\dpsi_b(\rvec,t) ,
\end{split}
\label{split}
\end{equation}
where the operator $\ha$ and $\hb$ obey bosonic commutation relations
$[\ha,\ha^\dagger]=1$ and $[\ha,\hb]=0$.  
In this paper we shall
derive the time evolution of the field operators by assuming that the
fluctuations $\dpsi_a$ and $\dpsi_b$  are much smaller than the
contributions from the condensate modes. An
approach similar to ours has previously been considered in 
Refs. \cite{castin-dum,gardiner,girardeau} where particle number conserving
Bogoliubov approximations are derived. We shall generalize these previous
results to a situation where the atoms can be in two different internal
states and where the number of atoms in each of the states may not be well
defined.  This generalization enables us to describe the collapse of
the relative phase between the two condensates and the entanglement of the
atoms, which are shown to arise from the time evolution of the
operators $\ha$ and $\hb$.

In section \ref{valid} we present the assumptions used in the 
Bogoliubov description \cite{castin-dum,gardiner} and in section
\ref{theory} we derive the equations of motion for the relevant
operators. In section \ref{solve} we solve these equation and in 
\ref{numerical} we apply our theory to the considered experimental setup
described in subsection \ref{setup}. 
Finally, we end the paper with a  conclusion  in section \ref{conclusion}.

\section{Assumptions}
\label{valid}
In this section we describe the assumptions of our theory,
which is presented in sections \ref{theory} and \ref{solve}. The
assumptions and methods used 
here are very 
similar to the two theories derived for a one component condensate by
Castin and Dum \cite{castin-dum} and by Gardiner \cite{gardiner}. These 
two derivations  are equivalent but use two different
approaches. Castin and Dum derive  equations of motion for the
relevant field operators and these equations are approximated by
inserting a splitting as in Eq.\ (\ref{split}) and neglecting high
order terms in $\dpsi$. In the work of Gardiner the approximations
are made to the Hamiltonian, where terms of high order in $\dpsi$ are
neglected before the equations of motion are derived.
Here we choose to follow the derivation and notation of Castin and Dum in
Ref. \cite{castin-dum} very closely. In order to make the paper self
contained  we reproduce most of the calculations of Castin and Dum.

We assume that the
atoms are cooled to a very low temperature such that a condensate is
formed  before the pulse which mixes the two internal states.    
The condensate wavefunction is determined from the one-body density operator
which is given by
\begin{equation}
\la \rvec'| \rho_{a,1}|\rvec\ra=\la \hpsi_a ^\dagger(\rvec)\hpsi_a
(\rvec')\ra. 
\end{equation}
Mathematically, the definition  of a condensate is that 
$\rho_{a,1}$ has an eigenvalue  
\begin{equation}
\rho_{a,1}|\Phi_{a} \ex\ra=N_{a}\ex |\Phi_a \ex \ra,
\label{eigen}
\end{equation}
which is much larger than all other
eigenvalues ($N_a\ex\sim N$, where $N$ is the total number of atoms),
and the spatial modefunction is given by $\Phi_a\ex (\rvec)$. 

The resonant
pulse mixes the field operators as described by Eq.~(\ref{pulse}), and by
using that the initial state is the vacuum of $\hpsi_b(t=0^-)$ we find that
$\rho_{a,1}(t=0^+)= \rho_{b,1}(t=0^+)=\rho_{a,1}(t=0^-)/2$. After  the
pulse the state $|\Phi_a\ex(t=0^-)\ra$ is still an eigenvalue of the
density operator $\rho_{a,1}$ with eigenvalue
$N_a\ex(t=0^+)=N_a\ex(t=0^-)/2$, so that we have a component of the
condensate in both 
the $a$ and the $b$ state with wavefunctions
\begin{equation}
 \Phi_a\ex(\rvec,t=0^+)= \Phi_b\ex(\rvec,t=0^+)= \Phi_a\ex(\rvec,t=0^-).
\label{initwave}
\end{equation}
 
The condensate modefunctions $\Phi_a \ex $ and $\Phi_b\ex$ can be used
to define a splitting 
of the field operator as in Eq.~(\ref{split}), and
to describe the squeezing we need to determine the time evolution of
$\dpsi_a$ and $\ha$. 
To derive the equations of motion for these operators it
will be convenient to have an expression for them in terms of
the field operator $\hpsi_a$.  By normalizing $|\Phi\ex_a\ra$ to
unity we obtain
\begin{equation}
\ha=\int d^3 r {\Phi_a\ex} ^*(\rvec)\hpsi_a(\rvec)
\label{haproj}
\end{equation}
and 
\begin{equation}
\dpsi_a(r)=\int d^3 r' \la \rvec|Q_a\ex|\rvec'\ra \hpsi_a(\rvec'),
\label{dpsiproj}
\end{equation}
where $Q_a\ex$ is the projector on the subspace perpendicular to
$|\Phi_a\ex\ra$: $Q_a\ex=1-|\Phi_a\ex\ra\la\Phi_a\ex|$. 
To simplify the notation we introduce the operand $\circ$ to denote
the action of an operator $O$ onto a field 
operator such as $\hpsi_a$:
\begin{equation}
O\circ \hpsi_a= \int d^3 r O|\rvec\ra \hpsi_a(\rvec)
\label{circ}
\end{equation} 
($\circ$ is also used in Ref.\ \cite{castin-dum}).
With this notation Eqs.~(\ref{haproj}) and (\ref{dpsiproj}) can be written
in the simpler form
\begin{equation}
\begin{split}
\ha&=\la\Phi_a\ex|\circ \hpsi_a  \\
\dpsi_a&=Q_a\ex\circ\hpsi_a.
\end{split}
\end{equation}
%(Note that we do not explicitly specify the dependence of $\rvec$ in
%$\dpsi_a$ and $Q_a\ex$).

Our calculation assumes a condensate of weakly interacting
atoms ($a_s^3\rho\ll 1$, where $\rho$ is the atomic density) at
very low temperature $T\approx 0$. In this limit the fraction of
non-condensed atoms is very small, $\delta N/N\sim 10^{-2}-10^{-3}$
\cite{lille}.  
The field operator $\hpsi_{a}$ has matrix elements scaling as $\sqrt{\bna}$
from 
the condensed atoms and contributions of order $\sqrt{\delta N_a}$ from the
non-condensed atoms, where $\bna$ and $\delta N_a$ are the average number of
particles and the 
number of uncondensed particles of
type $a$. Due to the small ratio between non-condensed and
condensed particles it is useful to perform an expansion of the field
operator $\hpsi_{a}$ in terms of $\sqrt{\delta N_a/\bna}$. Formally this
expansion is 
achieved by taking the limit
\begin{equation}
\begin{split}
\bna&\rightarrow \infty  \\
\bna g_{aa} &= {\rm constant},
\end{split}
\label{limit1}
\end{equation}
while keeping a constant ratio between the scattering lengths and between
the number of particles in the $a$ and the $b$ states.
Below we
show that $\delta N_a$ is of order unity in this limit, and our
expansion in terms of $\sqrt{\delta 
  N_a/\bna}$ becomes  
an expansion in $1/\sqrt{\bna}$  (note that since we have a fixed ratio
$\bna/\bnb$ this also corresponds to an expansion in
$1/\sqrt{\bnb}$).
% Physically, without changing the scattering lengths the
% limit 
% in Eq.~(\ref{limit1}) can be achieved by taking the limit 
% \begin{eqnarray}
%  \bna &\rightarrow& \infty \nonumber \\
% \frac{\bna a_{s,aa}}{l_0}&=&{\rm constant} \nonumber  \\
% \frac{k_b T}{\hbar \omega}&=&{\rm constant}, 
% \label{limit2}
% \end{eqnarray}
% where  $l_0=\sqrt{\hbar/M/\omega}$ is the width of the ground state for an
% isotropic harmonic trap 
% with frequency $\omega$. 

\section{Expansion in powers of $1/\sqrt{\bna}$}
\label{theory}

In this section we present the explicit expansion in powers of
$1/\sqrt{\bna}$. 
The expansion procedure is implemented by considering
the time evolution of the operator \cite{replaceN}
\begin{equation}
\hlam_a\ex (\rvec,t)=\frac{1}{\sqrt{\bna}}{a}^\dagger(t)\dpsi_a(\rvec,t).
\label{lambda}
\end{equation} 
From the definition of the condensate wavefunction (\ref{eigen}) and the
splitting (\ref{split}) follows that the expectation value of
$\hlam_a\ex$ vanish exactly at all times 
\begin{equation}
\langle \hlam_a\ex(\rvec,t)\ra=0.
\label{explambda}
\end{equation}

As we shall show below, $\hlam_a\ex$ is of order unity in the limit
in Eq.\ (\ref{limit1}) so that the number of non-condensed atoms approaches a
constant in this limit.  To perform our expansion we write the
operator $\hlam_a\ex$ and the wavefunction as a series of terms  
\begin{equation}
\begin{split}
\hlam_a\ex&=\hlam_a+\frac{1}{\sqrt{\bna}} \hlam_a^{(1)}
  +\frac{1}{\bna}\hlam_a^{(2)} + ....  \\
\Phi_a\ex&= \Phi_a+\frac{1}{\sqrt{\bna}}
  \Phi_a^{(1)}+\frac{1}{\bna} \Phi_a^{(2)}+... ,
\end{split}
\label{expansion}
\end{equation}
and we derive equations for each of the terms order by order. 
Since we require that the exact wavefunction is normalized to all orders in 
$1/\sqrt{\bna}$, the lowest order contribution must also be 
normalized $\la\Phi_a|\Phi_a\ra=1$. 

The procedure in this section is to calculate $d \hlam_a\ex/dt$ and
$d\ha/dt$ to order
$k$ in $1/\sqrt{\bna}$ ($k=-1,0$) and take the mean value $\la
\hlam_a\ex\ra$ which must vanish to all orders in
$1/\sqrt{\bna}$. With this procedure we calculate the contributions to the
field operator to order $\bna^0$. 
The main difference compared to the derivation of Castin and Dum in
Ref.\ \cite{castin-dum} is that 
in their derivation the number of atoms of type $a$ is fixed. This means that
$\hna-\bna$, where $\hna=\ha^\dagger\ha$ is the number operator for
the condensed particles, is only of order 
$\bna^0\sim 1$ because the difference only arises  due to the
non-condensed particles. In our situation the distribution on 
different number states is a binomial distribution with a width
$\sim\sqrt{\bna}$ and hence 
the operator $\hna-\bna$ is of order $\sqrt{\bna}$.

\subsection{Basic equations for $\hlam_a\ex$ and $\ha$}

As a starting point for our calculation we use the time derivatives of  the
operators $\hlam_a\ex$ and $\ha$.
By taking the time derivative of Eqs.~(\ref{haproj}) and (\ref{dpsiproj})
and using the definition of 
$\hlam_a\ex$ (\ref{lambda}) we obtain the expressions
\begin{equation}
i\frac{d}{dt} \ha =i {\left[{\left( \frac{d}{dt} \la\Phi_a\ex|\right)}\circ
  \hpsi_a +   \la\Phi_a\ex| \circ \ddt
  \hpsi_a \right]}
\label{dadt}
\end{equation}
and 
\begin{eqnarray}
i\ddt\hlam_a\ex& =& \frac{i}{\sqrt{\bna}} \bigglb[ {\ha}^\dagger
Q_a\ex \circ {\left( \ddt\hpsi_a-\ha \ddt|\Phi_a\ex\ra \right)}
  \nonumber \\ 
&& + {\left(  {\left(\ddt
        \hpsi_a^\dagger\right)}\circ
    |\Phi_a\ex\ra +\hpsi_a^\dagger\circ\ddt |\Phi_a\ex\ra \right)}
    \dpsi_a  \nonumber \\
&& - {\ha}^\dagger |\Phi_a\ex\ra
{\left(\ddt\la\Phi_a\ex| \right)}\circ \dpsi_a \biggrb] .
\label{dlamdt}
\end{eqnarray}
In 
these expressions the only time derivatives are of the wavefunction
$\Phi_a\ex$ and the field operator $\hpsi_a$. By substituting  
the splitting of the field operators in  Eq.~(\ref{split}) into the right
hand side of the time derivative of $\hpsi_a$ in Eq.~(\ref{dpsidt}) and
inserting this into the above expressions we may find equations of motion
for $\hlam_a\ex$ and $\ha$ involving different powers of $\dpsi_a$. Below
we shall only 
consider the two lowest order contributions to $\hlam_a\ex$ and $\ha$,
 and it is
sufficient to keep the terms which are linear in
$\dpsi_a$. Equivalently, this corresponds to neglecting the terms of
higher than second order in the $\dpsi$ operators in the Hamiltonian
if we had  used the approach of Gardiner \cite{gardiner}, where the
approximation are made to the Hamiltonian rather than in the equations
of motion. 
  
\subsection{Order $\sqrt{\bna}$: Gross-Pitaevskii equations}
The largest terms in the expression in Eq.~(\ref{dlamdt}) are the the terms
on the first line. These two terms are of order $\sqrt{\bna}$, but
with a suitable definition of the wavefunction 
$|\Phi_a\ex\ra$ 
the two contributions cancel each other to leading order in
$\sqrt{\bna}$, and $\hlam_a\ex$ is of order $\bna^0$. By using
Eqs.~(\ref{dpsidt})  and (\ref{split})
and keeping only the dominant contribution from the condensed particles we
get
\begin{equation}
i\ddt\hlam_a\ex={\sqrt{\bna}}Q_a\circ {\left( H_{{\rm GP},a} -i\ddt\right)}
 |\Phi_a\ra+O(\bna^0),
\label{dlamdt1}
\end{equation}
where we have introduced the Gross-Pitaevskii Hamiltonian
\begin{equation}
H_{{\rm GP},a}= H_{0,a}+g_{aa}\bna|\Phi_a|^2+g_{ab}\bnb|\Phi_b|^2.
\label{hgp}
\end{equation}
In Eq.~(\ref{dlamdt1}) we have replaced the number operators by their
leading order contributions $\ha^\dagger\ha \approx\bna$ and
$\ha^\dagger\ha^\dagger \ha\ha\approx \bna^2$. The difference between the
leading order and the operators is of higher order and will be taken into
account below. From Eq.~(\ref{dlamdt1}) follows that the $\sqrt{\bna}$
contribution to $\hlam_a\ex$ vanishes if we chose 
\begin{equation}
{\left( -i\ddt+H_{{\rm GP},a}\right)}|\Phi_a\ra=\zeta(t) |\Phi_a\ra.
\label{gpeq}
\end{equation}
The term on the right side introduces an optional phase factor on the
wavefunction $|\Phi_a\ra$ ($\zeta$ must be real due to the normalization
condition).  Since all measurable quantities are expressed in
terms of the field operator (\ref{split}), we can introduce any phase factor
on $|\Phi_a\ra$ if we include the opposite phase factor on $\ha$. Here
shall choose the simplest possible phase evolution of $\ha$ and we set
$\zeta=0$ so that (\ref{gpeq}) reduces to the usual time dependent
Gross-Pitaevskii 
equation. With this choice of $\zeta$ the equation of motion for $\ha$
cancels to leading order
$i d \ha/dt= O(\bna^0)$.
  
\subsection{Order $\bna^0$: $\Phi_a^{(1)}=0$.}
In this subsection we shall show that the lowest order correction to the
wavefunction $\Phi_a^{(1)}$ vanishes exactly in the
considered experimental situation. To show this, we consider the mean value
of Eq.~(\ref{dlamdt}) to order  $\bna^0$
\begin{equation}
\begin{split}
i\ddt \la \hlam_a(\rvec)\ra =&\sqrt{\bna}\la \rvec | Q_a\ex\circ \bigglb(
 H_{0,a}+g_{aa}\bna 
    |\Phi_a \ex|^2  \\
   & +g_{ab}\bnb|\Phi_b\ex|^2
   -i\ddt\biggrb) |\Phi_a\ex\ra 
   +O(1/\sqrt{\bna}). 
\end{split}
\label{meanlambda}
\end{equation}
When calculating the mean value of the third term in
Eq.~(\ref{dlamdt}) $\la d\hpsi_a^\dagger/dt ~ \dpsi_a\ra/\sqrt{\bna}$ we
need to consider  mean values like $g_{aa}\la \ha\ha^\dagger
\ha^\dagger\dpsi_a\ra/\sqrt{\bna}$. These terms do not vanish
and at first sight they might seem to contribute to the present
order. But since $\la \hlam_a\ex\ra$ vanishes, this term may be rewritten as
$g_{aa}\la (\hna-\bna) \ha^\dagger\dpsi_a\ra/\sqrt{\bna}$, and
since the parenthesis is only of order $\sqrt{\bna}$ this term does not
contribute to the present order.

Because of the exact
relation (\ref{explambda}) the 
mean value of Eq.~(\ref{dlamdt}) must vanish to all orders in
$\sqrt{\bna}$. The $\sqrt{\bna}$ contribution vanish due to the
Gross-Pitaevskii equation (\ref{gpeq}). The next order is obtained by
expanding $Q_a\ex$ and $\Phi_a\ex$ in Eq.~(\ref{meanlambda}) and we get 
\begin{equation}
\begin{split}
0=Q_a \circ&\bigglb[{\left(-i\ddt+ H_{0,a}\right)}\Phi_a^{(1)} \\
  & +g_{aa}\bna{\left(2|\Phi_a|^2 
    \Phi_a^{(1)} +\Phi_a^2{\Phi_a^{(1)}}^*\right)} \\
&+g_{ab}\bnb{\left(
    \Phi_a\Phi_b{\Phi_b^{(1)}}^*+\Phi_a\Phi_b^*\Phi_b^{(1)}
    +|\Phi_b|^2\Phi_a^{(1)} \right)}\biggrb]. \\
\end{split}
\end{equation}
Due to the projection operator $Q_a$ this equation admits the inclusion of
a an overall phase in $\Phi_a^{(1)}$, i.e., $i d \Phi_a^{(1)}/dt$ may
have a contribution  $ \zeta^{(1)} \Phi_a$.  Again we 
prefer to have the simplest possible  equation for $\ha$ and chose
$\zeta^{(1)}=0$. With this choice the equation for $\Phi_a^{(1)}$ is linear
and homogeneous so that $\Phi_a^{(1)}$ vanishes if it vanishes at
$t=0^+$. In Ref. \cite{castin-dum} it is shown that $\Phi_a^{(1)}=0$ in
thermal equilibrium as we have just before the resonant pulse
(\ref{pulse}), and since the condensate wavefunction is exactly the same
before and after the pulse, as described in Eq.~(\ref{initwave}), we have 
\begin{equation}
 \Phi_a^{(1)}(t)=0
\label{phi1}
\end{equation}
for all $t$.

\subsection{Order $\bna^0$: Bogoliubov approximation}
With the result (\ref{phi1}) we can now calculate the derivative in
Eq.~(\ref{dadt})  to order $\bna^0$ and we get
\begin{equation}
\begin{split}
i\ddt \ha =& g_{aa} \la \Phi_a| |\Phi_a|^2 |\Phi_a\ra
    (\hna-\bna )\ha \label{dadt2} \\
& +g_{ab} \la \Phi_a| |\Phi_b|^2
    |\Phi_a\ra 
    (\hnb-\bnb)\ha \\
 & +g_{aa} \sqrt{\bna}{\left( \la \Phi_a | |\Phi_a|^2\circ \hlam_a+
   \hlam_a^\dagger \circ 
    |\Phi_a |^2 
    |\Phi_a\ra  \right)}\ha   \\
  & +g_{ab}\sqrt{\bnb}{\left( \la \Phi_a | \Phi_b^* \Phi_a\circ \hlam_b+
    \hlam_b^\dagger \circ \Phi_a^* 
    \Phi_b|\Phi_a\ra  \right)}\ha,  
\end{split}
\end{equation}
where we have replaced expressions like
$\ha^\dagger\ha^\dagger\ha\ha-\ha^\dagger\ha\bna$ with their lowest order
contributions $(\hna-\bna)\bna$. The right hand side is seen to
be a Hermitian operator multiplied by $\ha$. This does not
contribute to the time derivative of 
the number operator $\hna$, so that we have
\begin{equation}
\ddt \hna= O(\bna^0).
\label{dndt}
\end{equation}
This relation is consistent with  $\hlam_a$ being of
order unity. The number of particles of type $a$ is a conserved quantity and
hence $\hna+\dpsi_a^\dagger\circ\dpsi_a=\hna
+\hlam_a^\dagger\circ\hlam_a+O(1/\sqrt{\bna})$ must be 
conserved. Since $\hlam_a$ is of order $\bna^0$, we must have a relation like
(\ref{dndt}). The weak time dependence of the number of condensed particles
indicate that essentially all particles stay in the condensate modes and
this justifies the description of the system by single mode
approximations. Note, however that such single mode calculations must be
performed with care. A calculation which completely ignores the
non-condensed particles only contains the first two lines in
Eq.~(\ref{dadt2}), but the remaining terms are of the same order as the
first and should be kept in the calculation. As we shall see below these
terms gives rise to the same interaction as in the single mode
approximation but with a different coupling constant. 

For the $\hlam$s the equation of motion reads
\begin{equation}
i\ddt{\left[
\begin{array}{c}
\hlam_a\\
\hlam_b\\
\hlam_a^\dagger\\
\hlam_b^\dagger
\end{array}
\right]}
={\mathcal L} \circ 
{\left[
\begin{array}{c}
\hlam_a\\
\hlam_b\\
\hlam_a^\dagger\\
\hlam_b^\dagger
\end{array}
\right]}
+(\hna-\bna)\vec{\alpha} 
+(\hnb-\bnb) \vec{\beta},
\label{dlamdt2}
\end{equation}
where the vectors $\vec{\alpha}$ and $\vec{\beta}$ are given by
\begin{equation}
\vec{\alpha}= |\Phi_a|^2
{\left[
\begin{array}{c}
g_{aa}\sqrt{\bna} |\Phi_a\ra\\
g_{ab}\sqrt{\bnb} |\Phi_b\ra\\
-g_{aa}\sqrt{\bna}| \Phi_a^*\ra\\
-g_{ab}\sqrt{\bnb}| \Phi_b^*\ra
\end{array}\right]}
\label{alpha}
\end{equation}
and 
\begin{equation}
\vec{\beta}=
 |\Phi_b|^2
{\left[
\begin{array}{c}
g_{ab}\sqrt{\bna}| \Phi_a\ra\\
g_{bb}\sqrt{\bnb}| \Phi_b\ra\\
-g_{ab}\sqrt{\bna}| \Phi_a^*\ra\\
-g_{bb}\sqrt{\bnb}| \Phi_b^*\ra
\end{array}\right]}
\label{beta}
\end{equation}
and where the matrix ${\mathcal L}$ is  
\begin{widetext}
\begin{equation}
{\left[
\begin{array}{cccc}
H_{{\rm GP},a}+Q_a g_{aa}\bna |\Phi_a|^2 Q_a &
Q_ag_{ab}\sqrt{\bna\bnb} \Phi_a\Phi_b^* Q_b & 
Q_a g_{aa}\bna \Phi_a^2 Q_a^* &
Q_a g_{ab}\sqrt{\bna\bnb}\Phi_a\Phi_b Q_b^* \\
%-----------------newline--------------------
Q_bg_{ab}\sqrt{\bna\bnb}\Phi_a^* \Phi_b Q_a & 
H_{{\rm GP},b}+Q_b g_{bb}\bnb |\Phi_b|^2 Q_b &
Q_b g_{ab}\sqrt{\bna\bnb}\Phi_a\Phi_b Q_a^* &
Q_b g_{bb}\bnb \Phi_b^2 Q_b^*\\
%-----------------newline-------------------- 
-Q_a^* g_{aa}\bna {\Phi_a^*}^2 Q_a &
-Q_a^* g_{ab}\sqrt{\bna\bnb}\Phi_a^*\Phi_b^* Q_b &
-H_{{\rm GP},a}-Q_a^* g_{aa}\bna |\Phi_a|^2 Q_a^* &
-Q_a^*g_{ab}\sqrt{\bna\bnb} \Phi_a^*\Phi_b Q_b^* \\
%-----------------newline-------------------- 
-Q_b^* g_{ab}\sqrt{\bna\bnb}\Phi_a^*\Phi_b^* Q_a &
-Q_b^* g_{bb}\bnb {\Phi_b^*}^2 Q_b&
-Q_a^* g_{ab}\sqrt{\bna\bnb} \Phi_a\Phi_b^* Q_b^* &
-H_{{\rm GP},b}-Q_b^* g_{bb}\bnb |\Phi_b|^2 Q_b^* 
\end{array}
\right]}.
\label{l}
\end{equation}
\end{widetext} 
Here we have introduced the projector $Q_a^*$ which projects onto the
space orthogonal to $|\Phi_a^*\ra$: $Q_a^*=1-|\Phi_a^*\ra
\la\Phi_a^*|$ (note that $Q_a^*\neq Q_a^\dagger=Q_a$).
The first part of Eq.~(\ref{dlamdt2}) containing the operator ${\mathcal
L}$ corresponds to the equation derived for a single component in
Ref.~\cite{castin-dum}. With a fixed
number of particles in the $a$ and the 
$b$ states this is the only contribution to this order. In a situation where
we cannot ignore the fluctuations in the particle numbers there are
additional contributions due to the last two terms in
Eq.~(\ref{dlamdt2}). Physically, we can understand the origin of these
terms by
considering a subspace with a given number of particles ${\cal N}_a$ and
${\cal N}_b$ (of type $a$ and $b$ respectively). In this subspace the
Gross-Pitaevskii equations are not the best 
approximations  
to the evolution of the condensate mode because they contain the average
values $\bna$ and $\bnb$ rather than ${\cal N}_a$ and
${\cal N}_b$. From the structure of the last two
terms in Eq.~(\ref{dlamdt2}) it is seen that these terms corrects for the
``wrong'' value of the  particle numbers. The 
approximate method used in Refs.\ \cite{anders-nature,sinatra} is
designed capture this dependence on the particle numbers and in section
\ref{numerical} we show that the results obtained with the present
method are very similar to the result obtained in Refs.\
\cite{anders-nature,sinatra}.

\section{Solving the equations}
\label{solve}

With the equations (\ref{dadt2}) and (\ref{dlamdt2}) we have derived
the equations of motion to the desired accuracy, and in this section we solve
these equations.  The solution is obtained by
expanding the $\hlam$ operators on a suitable set
of modefunctions at $t=0^+$. The  time evolution of these modes 
is chosen such that the first term in Eq.~(\ref{dlamdt2}) is automatically
taken into account and we then treat the last two terms in
Eq.~(\ref{dlamdt2}). Finally the solution for the $\hlam$ operators is
inserted into Eq.~(\ref{dadt2}) which is then solved. 

\subsection{Expansion at $t=0^+$}
Instead of having the operators $\hlam_a$
and $\hlam_b$ which have both a spatial part and an operator character, it is
convenient to expand $\hlam_a$ and $\hlam_b$ on a set of vectors which take
the spatial dependence into account. In such an expansion, the expansion
coefficients become operators and we have
\begin{equation}
{\left[
\begin{array}{c}
\hlam_a\\
\hlam_b\\
\hlam_a^\dagger\\
\hlam_b^\dagger
\end{array}
\right]}=\sum_{k=1}^\infty \hc_k 
{\left[
\begin{array}{l}
|u_{ak}\ra\\
|u_{bk}\ra\\
|v_{ak}\ra\\
|v_{bk}\ra
\end{array}\right]}
+
\hc_k ^\dagger
{\left[
\begin{array}{l}
|v_{ak}^*\ra \\
|v_{bk}^*\ra \\
|u_{ak}^*\ra \\
|u_{bk}^*\ra
\end{array}\right]}.
\label{decomp}
\end{equation}
 The $u_a$s and the $v_a$s can
be any functions perpendicular to $\Phi_a$ and $\Phi_a^*$ respectively,
but the whole set must be chosen such that the entire space perpendicular
to the condensates modes can be spanned by the vectors. A convenient choice
of the functions is discussed in section \ref{numerical}.

The $u$ and $v$ functions play the same role as the modefunction in
ordinary Bogoliubov theory and it is an advantage to choose the  standard
normalization 
\begin{eqnarray}
\la u_{ak}|u_{ak'}\ra+\la u_{bk}|u_{bk'}\ra -\la v_{ak}|v_{ak'}\ra-\la
v_{bk}|v_{bk'}\ra&=&\delta_{kk'}\nonumber \\
\la v_{ak}^*|u_{ak'}\ra+\la v_{bk}^*|u_{bk'}\ra -\la u_{ak}^*|v_{ak'}\ra-\la
u_{bk}^*|v_{bk'}\ra&=&0.\nonumber\\
\label{orthog}
\end{eqnarray}
With this normalization we may express the expansion
operators $\hc_k$ in terms of the $\hlam$ operators
\begin{equation}
\hc_k=\la u_{ak}|\circ \hlam_a+\la u_{bk}|\circ \hlam_b-\la v_{ak}|\circ
\hlam_a^\dagger -\la v_{bk}|\circ \hlam_b^\dagger,
\label{ck}
\end{equation}
and from the commutation relation $[\hlam_a^\dagger,\hlam_a]=Q_a$ we find
that the $\hc_k$s obey bosonic commutation relations
$[\hc_k,\hc_{k'}^\dagger]=\delta_{kk'}$ and $[\hc_k,\hc_{k'}] = 0$.

\subsection{Time evolution}
To evolve the expansion (\ref{decomp}) we need to choose the time
evolution of the modefunctions. A convenient choice is 
\begin{equation}
i\ddt{\left[
\begin{array}{l}
|u_{ak}\ra \\
|u_{bk}\ra \\
|v_{ak}\ra \\
|v_{bk}\ra 
\end{array}
\right]}
={\mathcal L} \circ 
{\left[
\begin{array}{l}
|u_{ak}\ra \\
|u_{bk}\ra \\
|v_{ak}\ra \\
|v_{bk}\ra 
\end{array}
\right]}.
\label{dudt}
\end{equation}
With this choice of time evolution, the normalization (\ref{orthog}) is
conserved and so is the commutation relation of the $\hc_k$ operators.
Indeed, the time evolution of $\hc_{k}$ only comes from the
fluctuations in the number of particles and is given by
\begin{equation}
i\ddt \hc_k=(\hna-\bna)f_{ak}+(\hnb-\bnb)f_{bk},
%(\la u_{ak}|,\la u_{bk}|,-\la v_{ak}|,-\la
%             v_{bk}|)\cdot \\
%\qquad{\left[(\hna-\bna)\vec{\alpha}+(\hnb-\bnb)\vec{\beta}\right]}. 
\label{dcdt}
\end{equation}
where
\begin{equation}
\begin{split}
f_{ak}=&(\la u_{ak}|,\la u_{bk}|,-\la v_{ak}|,-\la
             v_{bk}|)\cdot\vec{\alpha} \\
f_{bk}=& (\la u_{ak}|,\la u_{bk}|,-\la v_{ak}|,-\la
             v_{bk}|)\cdot\vec{\beta}.
\end{split}
\label{f}  
\end{equation}
This equation is easily solved and by using that $\hna$ and $\hnb$ do not
depend on time to this order of approximation we find
\begin{equation}
\begin{split}
\hc_k(t)=\hc_k(t=0^+)&-i(\hna-\bna)F_{ak}(t)\\
  &-i(\hnb-\bnb)F_{bk}(t),
\end{split}
\label{ct}
\end{equation}
where the functions $F_{ak}$ and $F_{bk}$ are defined by
\begin{equation}
\begin{split}
F_{ak}(t)=& \int_0^t dt' f_{ak}(t')\\
F_{bk}(t)=& \int_0^t dt' f_{bk}(t').
\end{split}
\label{F}  
\end{equation}
Finally, by inserting Eqs.~(\ref{decomp}) and (\ref{ct}) into
Eq.~(\ref{dadt2}) we obtain
\begin{widetext}
\begin{equation}
\begin{split}
i\ddt\ha=&(\hna-\bna){\left(
 g_{aa}\la\Phi_a||\Phi_a|^2|\Phi_a\ra
%\nonumber \\
%&& 
+2\sum_{k=1}^\infty  {\rm Im}(f_{ak}^*F_{ak})\right)}\ha
\\ 
&+(\hnb-\bnb){\left( g_{ab}\la\Phi_a||\Phi_b|^2|\Phi_a\ra
% \nonumber \\
%&&
+2\sum_{k=1}^\infty
 {\rm Im}(f_{ak}^*F_{bk}) \right)} \ha 
+{\left(\sum_{k=1}^\infty
 f_{ak}^*(t)\hc_k(t=0^+)+f_{ak}(t)\hc_k^\dagger(t=0^+)\right)}\ha 
%\nonumber \\
% i\ddt\hb&=&(\hnb-\bnb){\left(
%  g_{bb}\la\Phi_b||\Phi_b|^2|\Phi_b\ra
% %\nonumber \\
% %&& 
% +\sum_{k=1}^\infty  s_{bk}G_k+s_{bk}^*G_k^*\right)}\hb
% \nonumber\\ 
% &&+(\hna-\bna){\left( g_{ab}\la\Phi_b||\Phi_a|^2|\Phi_b\ra
% % \nonumber \\
% %&&
% +\sum_{k=1}^\infty
%  s_{bk}F_k+s_{ak}^*F_k^* \right)} \hb 
% +{\left(\sum_{k=1}^\infty
% s_{bk}\hc_k(t=0^+)+s_{bk}^*\hc_k^\dagger(t=0^+)\right)}\hb
.  
\end{split}
\label{dadtfinal}
\end{equation}
\end{widetext}
The expression in Eq.~(\ref{dadtfinal}) has the solution
\begin{equation}
\ha(t)=\exp{\left[-i(\eta_a(t)+\hat{\Theta}_a(t))\right]}\ha(t=0^+) ,
%  \nonumber \\
%&\hb(t)=\exp{\left[-i(\eta_b(t)+\hat{\Theta}_b(t))\right]}\hb(t=0^+&),
\label{at}
\end{equation}
where the Hermitian operator $\hat{\Theta}_a(t)$ 
is  just the time integral 
from $0$ to $t$ 
of the three terms multiplying $\ha$ on right hand side of
Eq.~(\ref{dadtfinal}). The phase 
$\eta_a$ arises because the term involving the $\hc_k$ and
$\hc_k^\dagger$ operators does not commute with itself at different
times. By differentiating the expression in $(\ref{at})$ with respect to
time and using the Baker-Hausdorff relation a few times, we find that
(\ref{at}) is indeed a solution of (\ref{dadtfinal}) if the phase is
given by
\begin{equation}
\eta_a(t)=\sum_{k=1}^\infty \int_0^t dt'
  {\rm Im}{\left(F_{ak}(t')f_{ak}^*(t')\right)} .
\label{eta}
\end{equation}
A derivation of this phase and a geometrical interpretation of it can be
found in \cite{ion.lang}.
With the expression (\ref{at})
we have finished our
derivation of the evolution of the field operator. In the next section we
apply the developed theory to describe phase collapse and entanglement. 

\section{Application of the theory}
\label{numerical}
In this section we apply the theory developed in sections
\ref{theory} and \ref{solve} to describe the collapse of
the relative phase between the two condensates and the entanglement of the
atoms. 
For simplicity we shall not be as general as in the previous sections and 
we only consider a symmetric interaction which we describe below.

\subsection{Symmetric interaction}
We consider the symmetric situation where $g_{aa}=g_{bb}\neq
g_{ab}$, and where the first pulse (\ref{pulse}) is a
$\pi/2$ pulse such 
that $\bna=\bnb$. We assume a system with a fixed total number of
particles $N=2\bna$ so that we have 
\begin{equation}
\hna-\bna=-(\hnb-\bnb)=(\hna-\hnb)/2+O(\bna^0). 
\label{dna}
\end{equation}
We also assume that the trapping potentials for the two different internal
states are identical, spherically symmetric, and harmonic $V_a=V_b=1/2
m\omega^2r^2$.   
With this symmetric choice of interactions the condensate wavefunctions are
exactly  
the same for the two components of the condensate
$\Phi_a=\Phi_b=\Phi$. Furthermore the symmetry can also be exploited in the
Bogoliubov modes. We divide the sum over $k$ in Eq.~(\ref{decomp}) into a
sum over terms which are even under the exchange of $a$ and $b$ ($+$ modes)
and  terms which are odd under exchange of $a$ and $b$ ($-$ modes), i.e., we
have 
\begin{equation}
{\left[
\begin{array}{c}
|u_{ak}\ra\\
|u_{bk}\ra\\
|v_{ak}\ra\\
|v_{bk}\ra
\end{array}\right]}=
{\left[
\begin{array}{c}
|u_{k}^+\ra\\
|u_{k}^+\ra\\
|v_{k}^+\ra\\
|v_{k}^+\ra
\end{array}\right]} 
\ {\rm or} \
 {\left[
\begin{array}{c}
|u_{k}^-\ra\\
-|u_{k}^-\ra\\
|v_{k}^-\ra\\
-|v_{k}^-\ra
\end{array}\right]}. 
\label{evenodd}
\end{equation}
In the remainder of this article superscript $+$ and $-$ on operators and
functions will refer to
these even and odd modes. The symmetry of the modes is reflected in the
functions defined in section \ref{solve} which obey
\begin{equation}
\begin{split}
f_{bk}^\pm=& \pm f_{ak}^\pm \\
F_{bk}^\pm=& \pm F_{ak}^\pm .
\end{split}
\label{sym}
\end{equation}
From these relations we find that the two components have the same
phase 
\begin{equation}
\eta_a=\eta_b
\end{equation}
and from the second relation and Eq.~(\ref{dna})
follows that the 
operators describing 
the + modes are independent of time
\begin{equation}
\ddt \hc_k^+ =0.
\end{equation}

\subsection{Initial conditions}
To evaluate the time evolution of physical quantities we need to relate
the different 
operators after the resonant pulse (\ref{pulse}) to the 
similar operators before the pulse. The relation between the operators before
and after the pulse is completely describe by Eq.~(\ref{pulse})  and
below we  extract some result from the
general relations.

Before the resonant pulse (\ref{pulse}) we assume that a condensate is
formed in the $a$ state and that there are no particles in the $b$
state. This 
means that $\ha(t=0^-)$ is of order $\sqrt{N}$ whereas $\hb(t=0^-)$ is of
order 
unity so that we have from Eqs.~(\ref{pulse}) and (\ref{initwave})
\cite{dpsib} 
\begin{equation}
\begin{split} 
\ha(t=0^+)&=\frac{1}{\sqrt{2}} (\ha(0^-)-\hb(0^-)) \approx\ha(0^-)/ \sqrt{2}
 \\
\hb(t=0^+)&= \frac{1}{\sqrt{2}} (\ha(0^-)+\hb(0^-))\approx \ha(0^-)/\sqrt{2}.
\end{split}
\label{abinit}
\end{equation}
These approximations should be handled with care since they violate
both the commutation relation and unitarity. The sign $\approx$ in
Eq.\ (\ref{abinit}) means
that the operators have approximately the same matrix elements. 

The operators $\hc^+_k$ can be related to the operators at $t=0^-$ by using
Eq.~(\ref{ck}). Inserting the approximate relations in (\ref{abinit}) we
obtain
\begin{equation}
\begin{split}
\hc_k^+(t=0^+)=\sqrt{2} \bigglb(& \la u_k^+(0^+)  |\circ \hlam_a(0^-) \\
& - \la  v_k^+(0^+) |\circ
    \hlam_a^\dagger (0^-) \biggrb),
\end{split}
\label{cplusinittemp}
\end{equation}
where we have introduced a $\hlam$ operator before the $\pi/2$ pulse
$\hlam_a(0^-)=\ha^\dagger(0^-)/\sqrt{N}\dpsi_a(0^-)$ which is also used in
Ref.~\cite{castin-dum}. As the other $\hlam$ operators, this operator can
be expanded on a set of 
Bogoliubov modes
$\hlam_a(0^-)=\sum_k \hc_{ak}(0^-)|u_{ak}(0^-)\ra
+\hc_{ak}^\dagger(0^-)|v_{ak}(0^-)\ra$, where the mode functions obey
orthogonality relations similar to Eq.~(\ref{orthog}). In \cite{castin-dum}
it is shown that by a suitable choice of the modes $|u_{ak}(0^-)\ra$ and
$|v_{ak}(0^-)\ra$ the 
pseudo-particle operators $\hc_{ak}(0^-)$ are the Bogoliubov operators
describing excitations of the condensates, and the ground state is the
vacuum of these operators. The simplest relation between the
operators before and after the pulse is obtained by choosing
$|u_k^+(0^+)\ra=|u_{ak}(0^-)\ra/\sqrt{2}$ and
$|v_k^+(0^+)\ra=|v_{ak}(0^-)\ra/\sqrt{2}$. With these modes  the operators
describing 
quasi-particle excitations in the + modes are exactly the same as the
operators describing excitations before the pulse
\begin{equation}
\hc_k^+(t=0^+)=\hc_{ak}(t=0^-).
\label{cplusinit}
\end{equation}

To find the initial condition for the operators $\hc^-_k$ we again use
relation (\ref{ck}) and the approximation in Eq.~(\ref{abinit}) and we find
\cite{dpsib} 
\begin{equation}
\begin{split}
\hc_k^-(t=0^+)=-\frac{1}{\sqrt{\bna}}
\bigglb(& \ha^\dagger (0^-)\la u_k^-(0^+)|\circ \dpsi_b(0^-)\\ 
 &  - \ha (0^-)\la
    v_k^-(0^+)|\circ 
    \dpsi_b^\dagger (0^-)\biggrb).
\end{split}
\label{cminusinit}
\end{equation}
The simplest initial condition is obtained by choosing $|v_k^-(0^+)\ra=0$. With
this choice $\hc_k^-(0^+)$ is proportional to $\dpsi_b(0^-)$. Since the
initial state is the vacuum of  $\dpsi_b(0^-)$ the initial state for
$\hc_k^-$ is also the vacuum state independent of the
modefunctions  $|u_k^-(0^+)\ra$ and the state of the condensate. 
In our numerical work described
below, we choose the  $u_k^-$s to be the excited states of the
Gross-Pitaevskii Hamiltonian before the pulse. 

\subsection{Phase collapse}
We now investigate the collapse of the relative phase between the two
components of the condensate. As a quantitative measure of the collapse we
use the reduction of the mean spin in the $x$-direction which is defined by
$\la J_x\ra ={\rm Re}
{\left(\la J_+\ra\right)}$, where $J_+$ is given in terms of the fields
operators by $J_+=
\hpsi _a^\dagger\circ \hpsi_b$. By inserting the splitting in
Eq.~(\ref{split}), using the approximation in Eq.~(\ref{abinit}), and
introducing the $\hlam$s we obtain 
\begin{equation}
\la J_+\ra=\la \ha^\dagger\hb\ra +
\la \hlam_a^\dagger\circ \hlam_b\ra +O(1/\sqrt{\bna}).
\label{jplus1}
\end{equation}
The first term in this expression represents the time evolution of the
condensed atoms and the second term is the contribution from the
non-condensed atoms. At short times the first term is of order $\bna$
whereas the second term is of order unity. However, as we shall see below,
the time scale of the collapse scales as $\sqrt{\bna}$, and since $d
\hlam_a/dt$ is of order unity, the second contribution may become comparable to
the first at the collapse time. We shall only work in the
dynamically stable region $g_{ab}<g_{aa}$. In the opposite case
$g_{ab}>g_{aa}$ it is energetically favourable for the two components to
separate so that any asymmetry in the two wavefunctions can grow with time
and the system is dynamically unstable \cite{sinatra}. By using the same
techniques as was used to show a similar result in \cite{castin-dum}, one
can show that equation of motion for the $u$s and $v$s (\ref{dudt})
corresponds to the time evolution of a perturbation of the condensate
wavefunction perpendicular to the wavefunction itself. Hence the
stability of the 
system is determined by the stability of the solution of the
Gross-Pitaevskii equation, and if the system is dynamically unstable the
number of non-condensed particles is expected to grow exponentially
with time, so 
that the approximations performed here are invalid. On the other hand,
in the region $g_{ab}<g_{aa}$ it is
energetically favourable for the two components to be overlapping so that
the system is dynamically stable and the number of non-condensed particles
performs small oscillations in time. This behaviour is confirmed by direct
numerical 
integrations of the equations. In the stable region the contribution from
non-condensed particles remain much smaller than the contribution of the
condensed particles, and in the following we ignore the second term in
Eq.~(\ref{jplus1}).

When we insert the solution in Eq.\ (\ref{at}) into Eq.\ (\ref{jplus1}) we
see that the phase collapse 
arises from the difference of the angles
$\hat{\Theta}_a-\hat{\Theta}_b$. By using the relations (\ref{dna}) and 
(\ref{sym}), the difference in the angles may be expressed as 
\begin{equation}
\hat{\Theta}_a-\hat{\Theta}_b=(\hna-\hnb)\lambda+4{\rm
  Re}{\left(\sum_{k=1}^\infty 
{F_{ak}^-}^*\hc_k^-(t=0^+)\right)} ,
\label{angledif} 
\end{equation}
where the function $\lambda$ is determined by
\begin{equation}
\ddt \lambda=(g_{aa}-g_{ab})\la \Phi| |\Phi|^2|\Phi\ra+4{\rm
  Im}{\left(\sum_{k=1}^\infty 
{f_{ak}^-}^*{F_{ak}^-}\right)} 
\label{chicorrect}
\end{equation}
and $\lambda(t=0^+)=0$.
The first term in the time evolution in Eq.~(\ref{angledif}) is the
same as the one obtained 
from the time evolution with a Hamiltonian like (\ref{hspin}).
% and the
%second term represents admixture 
%of noise from the Bogoliubov modes into the condensate modes. 
The coupling strength (\ref{chisimple}) was
derived by only considering 
 the contribution from the condensate modes, and this coupling
 constant is the same as the first term
in the correct coupling strength in
Eq.~(\ref{chicorrect}). The second term in Eq.\
(\ref{chicorrect}) 
represents a modification of the coupling constant due to the
Bogoliubov modes. 
The atoms in the condensate  mode does give the largest contribution to the
Hamiltonian 
(\ref{hamilton}), but we do not obtain the correct coupling constant
by only considering the terms from the condensate modes, because it is the
fluctuations of these terms, and not the terms them self,  which give
rise to phase  
collapse and squeezing. The fluctuations are of lower order and are
comparable to the 
contributions from the Bogoliubov modes, and the Bogoliubov modes have
to be included.
The
Hamiltonian (\ref{hspin}) is, however, still 
valid provided that we use the correct coupling constant in
Eq.~(\ref{chicorrect}). 
 
\begin{figure}[bt]
  \centering
  \epsfig{angle=270,width=\figwidth,file=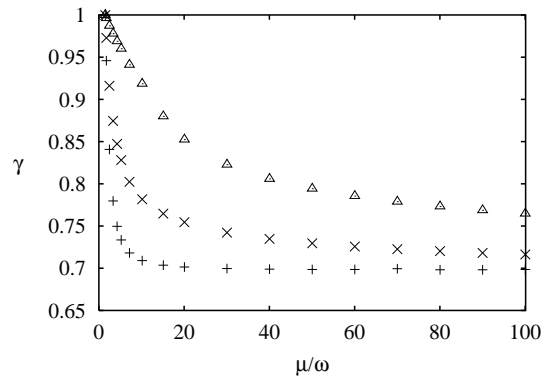}
  \caption[]{Modification of the coupling constant due to the Bogoliubov
  modes. The figure shows the ratio $\gamma$ between the average of
  the correct 
  coupling constant $d\lambda/dt$ (\ref{chicorrect}) and the average
  of the single mode value 
  (\ref{chisimple}) for $a_{s,ab}/a_{s,aa}=0$ ($+$),
  $a_{s,ab}/a_{s,aa}=0.5$ ($\times$),
  and $a_{s,ab}/a_{s,aa}=0.93$ ($\triangle$).
  The last value corresponds to the value for $|F=1,M_F=\pm 1\ra$ sodium
  atoms \cite{williams}.  
  In the limit of very weak interactions ($\mu\approx
  3/2 \omega$), $\gamma$ approaches the estimate in
  Eq.\ (\ref{chisimple}), 
  and in the 
  Thomas-Fermi limit ($\mu \gg \omega$) it approaches $7/10$ as predicted in
  Ref.\ \cite{sinatra}.}
  \label{fig:chi}
\end{figure}

The presence of the second term in Eq.~(\ref{chicorrect}) only changes the
coupling constant slightly. Ref.~\cite{sinatra} considers the same
situation (by a different method)
and in the 
Thomas-Fermi limit ($\mu/\omega\gg 1$, where $\mu$ is the chemical
potential) it is shown that the effective coupling constant is
approximately 7/10 of the coupling constant in Eq. (\ref{chisimple}). The
functions $f_{ak}^-$ and  $F_{ak}^-$ appearing in the second term in
Eq.~(\ref{chicorrect}) 
comes from the change in the condensate wavefunctions due to the
fluctuations in the number of particles in states $a$ and $b$. Outside the
the Thomas-Fermi regime the condensate wavefunctions are less affected by
the interactions and are closer to the ground state of the harmonic
potential which is independent of the number of particles of type $a$ and
$b$. We therefore expect that the contribution of the second term becomes
smaller if we go away from the Thomas-Fermi limit. This is indeed confirmed
by a numerical integration of the equations. In Fig.~\ref{fig:chi} we show
the ratio $\gamma$ between the slopes of  linear approximations to the
time integrals of
Eqs.~(\ref{chicorrect}) and (\ref{chisimple}). In the non-interacting limit
($\mu\approx 3 \omega/2$), Eq.~(\ref{chisimple}) gives the correct coupling
constant and in the Thomas-Fermi limit 
($\mu\gg\omega$) the correct coupling constant (\ref{chicorrect})
approaches 7/10 of 
Eq.~(\ref{chisimple}). Note, that this is only true for a time average of
the coupling strength. The expression in Eq.~(\ref{chicorrect}) have
larger oscillation in time. 

\begin{figure}[ht]
\centering
\epsfig{width=\figwidth,file=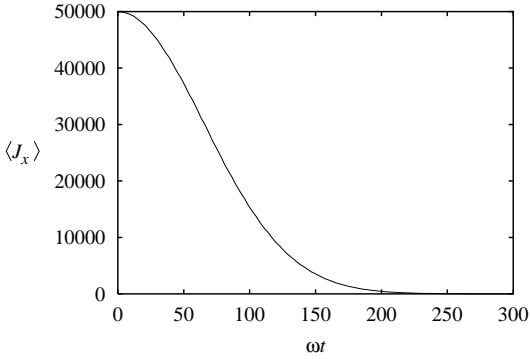}
\caption[]{Collapse of the relative phase between the two components
  of the condensate. The
  mean value of the spin $\la J_x \ra$ decreases due to the fluctuation in the
  number of particles in each of the internal states. The parameters are
  $a_{s,aa}/\sqrt{\hbar/M/\omega}=6\cdot 10^{-3}$,
  $a_{s,bb}=a_{s,aa}=2a_{s,ab}$, and $N=10^5$.} 
  \label{fig:collapse}
\end{figure}

If we ignore the non-condensed particles, the initial state is a
Fock state with $N$ particles of type $a$ in the condensate mode and the
vacuum of $\hb$ and $\hc_k^-$ for all $k$.
 By using the exact relations in
Eq.~(\ref{abinit}) we can calculate the shape of the phase collapse
\begin{equation}
\la J_x\ra =\frac{N}{2} \cos ^{N-1}(\lambda)\exp{\left(-2\sum_{k=1}^\infty
             |F_{ak}^-|       ^2 \right)}.
\label{jx}
\end{equation}
An example of the phase collapse is shown in Fig.~\ref{fig:collapse}, where
we show the phase collapse obtained by numerically integrating the
equations. The
functions $F_{ak}^-$ perform small oscillations in time and the last
exponential in Eq.~(\ref{jx}) is of minor importance. The phase collapse
arises from the cosine in Eq.~(\ref{jx}). The value of $\la J_x\ra$ is
reduced by a factor 
$1/e$ when $\lambda\approx \sqrt{2/(N-1)}$ and the collapse time is roughly
given by 
\begin{equation}
t_c=\sqrt{\frac{2}{N-1}} \frac{1}{\gamma\bar{\chi}},
\label{tcollapse}
\end{equation}
where $\bar{\chi}$ is the average of the coupling constant in
Eq.~(\ref{chisimple}). Since $\chi$ is proportional to $1/N$, $t_c$
scales as $\sqrt{N}$.

\subsection{Spin squeezing}
\label{squeezing}
We now turn to the  calculation of spin squeezing.
The calculation of  spin squeezing is more complicated than the
calculation of the phase collapse because squeezing is a lower order
phenomena. The
operators describing the noise are the square of the angular momentum
operators and involve the expectation values of the product of four
field operators. Such a product is in itself of order $N^2$ but due to
interferences between different terms of order $N^2$ the noise is only of
order $N$. Hence to calculate the squeezing we need to be very precise when
we calculate the matrix elements. Here we calculate all matrix elements to
order $N$ and omit terms of order $\sqrt{N}$ and smaller.

We are interested in estimating the optimal squeezing obtainable with the
proposal \cite{anders-nature}. Experimentally it is possible to
optimize the measurement output by choosing the angle $\theta$ such that
the variance is 
minimized, and by
minimizing the variance $(\Delta J_\theta)^2$ with respect to 
$\theta$ we find that the minimum variance is given by 
\begin{equation}
(\Delta J_\theta)^2_{{\rm min}}=\frac{\la J_z^2+J_y^2\ra-\sqrt{\la
  J_z^2-J_y^2\ra^2+\la J_zJ_y+J_yJ_z\ra^2}}{2}
\label{minvar}
\end{equation}
and that the optimal angle is determined by
\begin{equation}
\tan(2\theta)=\frac{\la J_zJ_y+J_yJz\ra}{\la J_z^2-J_y^2\ra}.
\label{angle}
\end{equation}

To calculate the variance $(\Delta J_\theta)_{\rm min}^2$ it is an
advantage to replace $J_y$ by the raising and lowering operators
$J_y=(J_+-J_-)/2i$, so 
that, e.g., $\la J_y^2 \ra=\la J_+ J_-\ra/2- {\rm Re}(\la J_+ J_+\ra)/2$,
where we have used the symmetry to obtain $\la J_+J_-\ra =\la J_-J_+\ra$. 
By expressing the raising and lowering operators in terms of the field
operators, using the splitting (\ref{split}), and introducing the $\hlam_a$
operator (\ref{lambda}) we obtain
\begin{equation}
\la J_+ J_-\ra =\la \ha^\dagger \ha \ra +\la \ha^\dagger\ha\hb^\dagger
\hb\ra +\bna(\la\hlam_b^\dagger \circ\hlam_a\ra+\la
\hlam_a^\dagger\circ \hlam_b \ra),
\label{jplusjminus1}
\end{equation}
where we have neglected terms of order $\sqrt{\bna}$. To calculate the
first two 
terms in this expression we use the fact that we have conservation of
the number of particles, and that $J_z$ commutes with the Hamiltonian
because we have no spin changing collisions. From the mean number of
particles in the $a$ state $\bna=\la \hpsi_a\circ\hpsi_a\ra$ we find
\begin{equation}
\la \ha^\dagger \ha \ra=\bna +O(\bna^0),
\end{equation}
from $\la J_z^2\ra= N/4=\la (\hpsi_a^\dagger \circ\hpsi_a-\bna)^2\ra $ we find 
\begin{equation}
\la (\ha^\dagger\ha)^2\ra= \frac{N(N+1)}{4}-N\la
\hlam_a^\dagger\circ \hlam_a\ra +O(\sqrt{\bna}),
\end{equation}
and with this expression and $N^2=\la(\hpsi_a^\dagger \circ \hpsi_a+
\hpsi_b^\dagger \circ \hpsi_b)^2\ra$ we get
\begin{equation}
\la \ha^\dagger \ha \hb^\dagger \hb\ra=
\frac{N(N-1)}{4}-N\la\hlam_a^\dagger \circ \hlam_a\ra+O(\sqrt{\bna}).
\end{equation}
In the last equation we have made a replacement $\la \dpsi_a^\dagger \hb
\hb^\dagger \dpsi_a\ra=N/2\la \hlam_a^\dagger \circ
\hlam_a\ra+O(\sqrt{\bna})$. This replacement is valid because $\hb\approx
\ha$  to lowest order according to Eq.~(\ref{abinit}). With
these equations we can calculate $\la J_+J_-\ra$ in (\ref{jplusjminus1})
and by doing a similar calculation for $\la J_+ J_+\ra$ we find
\begin{equation}
\la J_y^2 \ra = \frac{N(N+1)}{8}-\frac{N}{2}\la \hlam_a^\dagger
\circ\hlam_a\ra -\frac{1}{2}{\rm Re}(\la \ha^\dagger\ha^\dagger\hb\hb\ra).
\label{jysq1}
\end{equation}
By similar arguments we can also calculate 
\begin{equation}
\la J_zJ_y+J_yJ_z\ra =2 {\rm Im}(\la \ha^\dagger\ha^\dagger\ha\hb\ra).
\label{jzjy1}
\end{equation}

The matrix elements appearing in Eqs.~(\ref{jysq1}) and (\ref{jzjy1})
cannot be calculated to the desired accuracy by only using Eq.~(\ref{at}),
because this equation does not go to high enough accuracy. We need to
calculate the matrix elements to order $N$ but in Eq.~(\ref{at}) we are
omitting  contributions to $\ha$ of order $1/\sqrt{N}$. The next order
correction to $\ha$ gives a contribution of order $N$ when inserted in
Eqs.~(\ref{jysq1}) and (\ref{jzjy1}). However we can use the conservation of
the number of particles in each of the internal states to find the
remaining term in 
the matrix elements without going to higher order in the calculation of
$\ha$. We split the exact operator $\ha$  appearing in Eqs.~(\ref{jysq1})
and (\ref{jzjy1}) into the part that we have calculated so far $\ha_0$ and
an additional term $\delta\ha$ of order $1/\sqrt{N}$
\begin{equation}
\ha=\ha_0+\delta\ha.
\label{asplit}
\end{equation}
At $t=0^+$ we have $\delta\ha=0$ and by using that the number of particles in
state $a$ and $\ha_0^\dagger\ha_0$ are independent of time we find
\begin{equation}
\begin{split}
\ha_0^\dagger (t)\delta\ha(t)+\delta\ha^\dagger(t)\ha_0(t)=&
\hlam_a^\dagger(0^+)\circ 
\hlam_a 
(0^+) \\ &- \hlam_a^\dagger(t)\circ \hlam_a(t).
\end{split}
\end{equation}
Note, that we only need the lowest order contribution of this term, and
since $\ha(t)\approx\ha(0^+)\approx\hb(0^+)\approx\hb(t)$ according to
Eqs.\ (\ref{dadtfinal}) and (\ref{abinit}), we can actually
replace $\ha_0$ by $\hb_0$ which is necessary to complete the calculation
below. 

With the above relation we are finally able to calculate the
relevant matrix elements. Using the splitting in Eq.~(\ref{asplit}), the
initial condition in Eq.~(\ref{abinit}), the time evolution in
Eq.~(\ref{at}), and replacing $\ha(t=0^-)^\dagger\ha(t=0^-)$ with
$N-\hlam_a(t=0^-)^\dagger\hlam_a(t=0^-)=
N-2\hlam_a(t=0^+)^\dagger\hlam_a(t=0^+)$ we obtain
\begin{equation}
\begin{split}
\hspace{-0.3cm}\la J_y^2\ra =& \frac{N(N+1)}{8}  \\
&-\frac{N(N-1)}{8}
\cos^N(2\lambda)\exp {\left(-8\sum_{k=1}^\infty
             |F_{ak}^-|^2 \right)}
\end{split}
\label{jysq2}
\end{equation}
and 
\begin{equation}
\begin{split}
\la J_zJ_y+J_yJ_z\ra =-& \frac{N(N-1)}{4}\cos^N(\lambda)\sin(\lambda)
 \\ 
&\times\exp {\left(-2\sum_{k=1}^\infty
             |F_{ak}^-|^2 \right)}.
\end{split}
\label{jzjy2}
\end{equation}
By inserting these two expressions into Eq. (\ref{minvar}) we can find the
minimum variance and with the expression for $\la J_x \ra$ in Eq.\
(\ref{jx}) we may find the squeezing parameter $\xi_\theta^2$. 

Note
that we have not specified anything about the initial state of the $+$
modes. Within the approximations used here, the results (\ref{jysq2})
and  (\ref{jzjy2}) only involves expectation values of operators for
the $-$ modes and the squeezing is independent of the state of the $+$
modes. If the condensate is at a non-zero temperature before the
pulse, it can be described by quasi-particle excitations, and these
excitation are transfered into excitation in the $+$ modes after the
pulse. The $-$ modes  are always in the vacuum state independent of
temperature, and the squeezing calculated here is therefore also
independent 
of temperature. Another nice property of Eqs.\ (\ref{jysq2}) and
(\ref{jzjy2}) is that they only involve the functions $F_{ak}^-$ and
not the number of non-condensed particles. Because the functions $F_{ak}^-$
vanish for states with non-zero orbital angular momentum, according to
Eq.\ (\ref{f}),  it is
sufficient to determine the evolution of the Bogoliubov modes with
vanishing orbital angular momentum and this simplifies a numerical
treatment of the system significantly.

\begin{figure}
\centering
\epsfig{angle=270,width=\figwidth,file=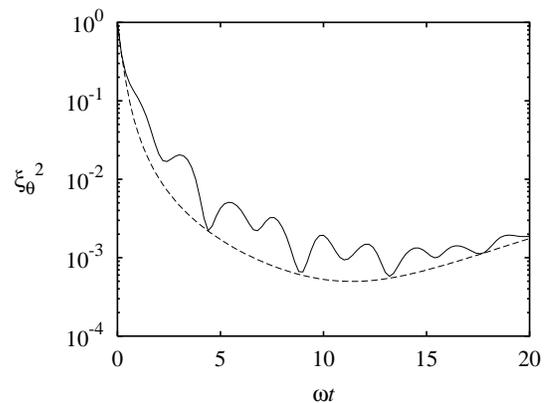}
\caption[]{Time Evolution of the squeezing calculated by a numerical
  integration (full line) and from the approximate Hamiltonian
  (\ref{hspin}) (dashed curve). The parameters are the same as in Fig.\
  \ref{fig:collapse}. } 
\label{fig:squeez05}
\end{figure}

In Fig.\
\ref{fig:squeez05} we show the evolution of spin squeezing with the same
parameters as in Fig.\ \ref{fig:collapse}. In the figure we also show
the prediction of the 
Hamiltonian (\ref{hspin}) with a coupling constant $\chi$ equal to the
slope of a linear approximation to $\lambda$. The two curves are in
reasonably good agreement with each other. The deviation between the two
curves is caused by the exponentials of the $F_{ak}^-$ functions in Eqs.\
(\ref{jysq2}) and (\ref{jzjy2}), i.e., by the dependence of the
wavefunction on the number of particles. After the resonant pulse the
wavefunction is no longer in equilibrium because the repulsion of the
atoms is suddenly reduced, and the size of the atomic cloud will
oscillate in time. The oscillations of the wavefunction depend on the
number of atoms in each of the internal state, and because the state of the
system is distributed on states with different number of atoms in each
internal state, there is an uncertainty in the wavefunction which introduces
noise and reduces the squeezing. After the completion of a full
oscillation ($\omega t\approx 4$, 9, 13, and 18), the wavefunction has
approximately its initial form independent of the number of atoms in each
internal state ($F_{ak}^-\approx 0$), and the results of the Hamiltonian
(\ref{hspin}) are in very good agreement with the numerical results.  
The result in Fig.\
\ref{fig:squeez05} is very similar to the result obtained with the
same parameters but a
different approximate method in Ref.\ \cite{anders-nature}, and our
result thus supports 
the conclusions reached in that paper. 

\begin{figure}
\centering
\epsfig{angle=270,width=\figwidth,file=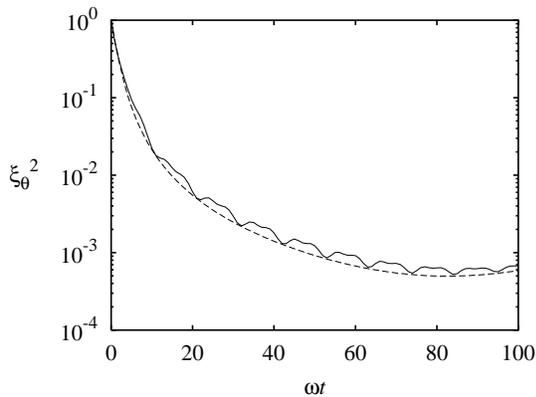}
\caption[]{Squeezing between the two $|F=1,M_F=\pm 1\ra$ hyperfine states in
  sodium. The parameters are the same as in Fig.\
  \ref{fig:collapse} except $a_{s,ab}/a_{s,aa}=0.93$. The full curve
  is the result of a numerical simulation and the dashed curve is from the
  Hamiltonian (\ref{hspin}).} 
\label{fig:na}
\end{figure}

The difference in scattering length used in Fig.\ \ref{fig:squeez05} is an
exaggeration of the realistic parameters. If the experiment is performed
with the two hyperfine states $|F=1, M_F=\pm 1 \ra$ in sodium representing
the internal states $a$ and $b$ as proposed in \cite{anders-nature},
the ratio between the scattering lengths is 
$a_{s,ab}/a_{s,aa}\approx 0.93$ \cite{williams}. In Fig.\
\ref{fig:na} we show the squeezing produced with this value of the
ratio. Because the scattering lengths are very close, the oscillations of the
wavefunction are much smaller, and the numerical
curve and the result from Eq.\ (\ref{hspin}) are in better agreement. 
% The
% ratio $a_{s,aa}/\sqrt{\hbar/M/\omega}=6 \cdot 10 ^{-3}$ used in the
% figure corresponds to a 
% trapping frequency 
% of $\omega\approx 2\pi \cdot 1.9$kHz.

To create an entangled state the
coupling constant $\chi$ in the Hamiltonian (\ref{hspin}) should be
non-zero. So far we have obtained a non-zero coupling by having a difference
in the the scattering lengths $a_{s,aa}+a_{s,bb}\neq2 a_{s,ab}$. From the
approximate coupling constant in Eq.\ 
(\ref{chisimple}) we see that $\chi$ is also non-zero if the $a$ and the
$b$ atoms occupy two different regions in space, even if all scattering
lengths are identical.
In Ref.\ \cite{uffe2} it was proposed to produce spin squeezing by
replacing the two internal states $a$ and $b$ with two different momentum
states which are created by 
a Bragg pulse shortly after the trap has been turned off, i.e., exactly the
same setup as used in \cite{phillips}. This proposal has the advantage that
the squeezing is now between atoms in the same internal state, and the
squeezing is therefore better shielded from phase decoherence caused for
instance by fluctuating magnetic fields. Also, this proposal could be used
for rubidium atoms where the scheme which has been studied so far is not
applicable because the scattering lengths are almost identical.

\begin{figure}[b]
  \centering
  \epsfig{angle=270,width=\figwidth,file=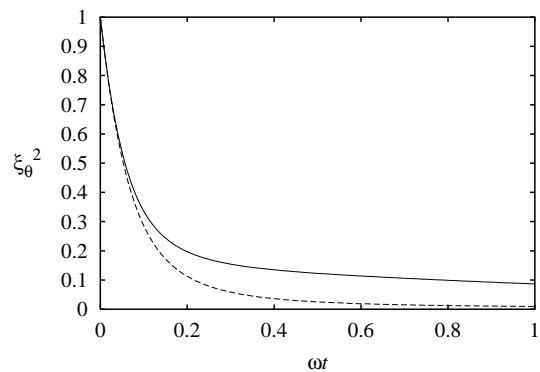}
\caption[]{Squeezing created by a Bragg pulse as proposed in Ref.\
  \cite{uffe2}. At time $t=0$ the trap is turned off and two
  components are separated by a Bragg pulse. The full line is
  the result of a numerical integration and the dashed curve is the
  prediction from the Hamiltonian (\ref{hspin}).
  The parameters are
  $a_{s,aa}/\sqrt{\hbar/M/\omega}=6\cdot 10^{-4}$,
  $a_{s,bb}=a_{s,aa}$, $a_{s,ab}=0$, and 
  $N=1.7 \cdot 10^6$. These parameters are chosen such that they
  are similar to the parameters in Ref.\ \cite{phillips}. The ratio
  $a_{s,aa}/\sqrt{\hbar/M/\omega}=6\cdot 10^{-4}$ corresponds to
  sodium atoms in  a trap 
  with trapping frequency 
  $\omega=2\pi\ \cdot 19$Hz before the trap is turned off.}
\label{fig:free}
\end{figure}

To investigate this situation we assume that the time it takes the two
momentum states to separate is very short so that we can neglect the
interaction during this separation process. In this case there is no
interaction between the $a$ and the $b$ components and we can describe this
situation by setting $a_{s,ab}=0$ and $V_a=V_b=0$ after the pulse. In Fig.\
\ref{fig:free} we show the result of such a simulation with parameters
similar to the parameters in Ref. \cite{phillips}. The
calculation shows that strong squeezing can be produced with this
proposal, but the agreement with the Hamiltonian (\ref{hspin}) is not
as good in this case  because of the exponentials in Eqs.\ (\ref{jysq2}) and
(\ref{jzjy2}). 

\section{Conclusion}
\label{conclusion}
In this paper we have analysed the scheme 
proposed in Ref.\ \cite{anders-nature} by describing the system with a
number-conserving Bogoliubov theory. The developed theory is a
consistent expansion in the ratio between non-condensed and condensed
particles, and the validity of the calculations can be investigated in
a given experimental configuration. 

The results obtained in this paper show that strong
squeezing can indeed be produced 
by this method, but the theory is not able to determine the precise limit
of the obtainable squeezing. The results of this paper agree with the
result of the simplified Hamiltonian (\ref{hspin}) which predict a
reduction of the squeezing parameter by a factor of approximately
$N^{-2/3}$, but we are ignoring terms of order $\sqrt{\delta N/N}$. From
this we conclude that the  obtainable squeezing is at least of
order  $\sqrt{\delta N/N}$.
The obtained results are independent of the temperature of the
condensate, and the only effect of a non-vanishing temperature is that 
the approximations break down a little earlier. 

\begin{acknowledgments}
I am grateful to Klaus M\o lmer and Uffe V.\ Poulsen for useful
discussions and comments on the manuscript.  
This work was supported by the Danish National Research Foundation
through the Quantum Optics Center (QUANTOP) in \AA rhus.
\end{acknowledgments}


\begin{thebibliography}{99}
\bibitem{replaceN} Compared to the derivation in Ref.~\cite{castin-dum} we
  have  replaced $\hat{N}_a$ appearing in the definition of $\hlam\ex_a$
  (\ref{lambda}) 
  with the number $\bna$. In our situation the number of particles in the
  $a$ state is  not a fixed number, and the arguments leading to
  Eq.~(\ref{explambda}) are easier to perform with a number $\bna$.
\bibitem{dpsib} Actually, $\hb$ and $\dpsi_b$ are not uniquely defined before
  the pulse 
  because there is no $b$ component of the condensate. The operators
  appearing 
  in Eqs. (\ref{abinit}) and (\ref{cminusinit}) are defined by using
  the condensate 
   modefunction for the 
  $a$ particles.
\end{thebibliography}
\end{document}